\documentclass{aa}  
\usepackage{graphicx}
\usepackage{xcolor}
\usepackage{txfonts}
\usepackage{hyperref}
\usepackage{graphicx}	
\usepackage{amsmath}	
\usepackage{amssymb}
\usepackage{tikz}
\usepackage{multirow}

\newcommand{\zgauss}{{\fontfamily{qcr}\selectfont zgauss}}
\newcommand{\pow}{{\fontfamily{qcr}\selectfont pl}}
\newcommand{\laor}{{\fontfamily{qcr}\selectfont laor}}
\newcommand{\krd}{{\fontfamily{qcr}\selectfont kerrdisk}}

\newcommand{\tbabs}{{\fontfamily{qcr}\selectfont tbabs}}

\newcommand{\ztbabs}{{\fontfamily{qcr}\selectfont ztbabs}}
\newcommand{\xillver}{{\fontfamily{qcr}\selectfont xillver}}

\newcommand{\apec}{{\fontfamily{qcr}\selectfont apec}}
\newcommand{\relxil}{{\fontfamily{qcr}\selectfont relxill}}

\usepackage{newtxtext,newtxmath}

\graphicspath{{./}{Figures/}}

\begin{document} 
 \title{Detection of the Fe K lines from the binary AGN in 4C+37.11}
   \subtitle{}
   \author{Santanu Mondal\inst{1}
    \and
         Mousumi Das \inst{1}   
    \and 
    K. Rubinur \inst{2}
    \and 
    Karishma Bansal \inst{3}
    \and
    Aniket Nath \inst{1,4}
    \and
    Greg B. Taylor \inst{5}
}
 \institute{Indian Institute of Astrophysics, 2nd Block Koramangala, Bangalore 560034, Karnataka, India\\
              \email{santanumondal.work@gmail.com, santanu.mondal@iiap.res.in}
           \and
           Institute of Theoretical Astrophysics, University of Oslo, P.O box 1029 Blindern, 0315 OSLO, Norway
           \and 
           National Radio Astronomy Observatory, Socorro, NM 87801
           \and 
           School of Physical Sciences, National Institute of Science Education and Research, An OCC of Homi Bhabha National Institute, Jatni-752050, India
           \and
           Department of Physics and Astronomy, University of New Mexico, Albuquerque, NM 87131
           }

  \abstract
  {We report the discovery of the Fe K line emission at $\sim6.62^{+0.06}_{-0.06}$ keV with a width of $\sim0.19^{+0.05}_{-0.05}$ keV using two epochs of {\it Chandra} archival data from the nucleus of the galaxy 4C+37.11, which is known to host a binary supermassive black hole (BSMBH) system where the SMBHs are separated by $\sim7$ mas or $\sim$ 7pc. Our study reports the first detection of the Fe K line from a known binary AGN, and has an F-statistic value of 20.98 and probability $2.47\times 10^{-12}$. Stacking of two spectra reveals another Fe K line component at $\sim7.87^{+0.19}_{-0.09}$ keV. Different model scenarios indicate that the lines originate from the combined effects of accretion disk emission and circumnuclear collisionally ionized medium. The observed low column density favors the gas-poor merger scenario, where the high temperature of the hot ionized medium may be associated with the shocked gas in the binary merger and not with star formation activity. The estimated total BSMBH mass and disk inclination are $\sim1.5\times10^{10}$ M$_\odot$ and $\gtrsim75^\circ$, indicating that the BSMBH is probably a high inclination system. The spin parameter could not be tightly constrained from the present data sets.  Our results draw attention to the fact that detecting the Fe K line emissions from BSMBHs is important for estimating the individual SMBH masses, and the spins of the binary SMBHs, as well as exploring their emission regions.}


   \maketitle



\section{Introduction} \label{sec:intro}
Binary Supermassive black holes (BSMBHs) are rare systems that may form during galaxy mergers \citep{BegelmanEtal1980Natur.287..307B}. When both the SMBHs are accreting mass they become active galactic nuclei (AGN) and depending on their separations are referred to as binary AGN ($<100$ pc), or dual AGN (DAGN; $>10$ kpc) \citep{Burke-Spolaor2014arXiv1402.0548B, rubinur2019}. Although there have been several detections of dual AGN, the detection of binary AGN is more challenging as it requires very high-resolution observations over different energy bands (e.g. see a recent review by \citet{DeRosaEtalReview2019NewAR..8601525D}). These systems are important for understanding the final stages of galaxy mergers, especially when the SMBHs are separated by a few parsecs, as then they can become gravitationally bound. When this happens, the binary SMBH system can emit gravitational waves that will contribute to the gravitational wave background \citep{agazie.etal.2023}. Several surveys in multi-wavelength bands are focused on searching for DAGN and binary AGN \citep[][and references therein]{AbazajianSDSS..Survey2009ApJS..182..543A,WrightWISE..Survey2010AJ....140.1868W,AhnBOSS..SDSS..Survey2012ApJS..203...21A,LenaEtal2018MNRAS.478.1326L}. 

\begin{figure*}
\centering{
\hspace{-6.0cm}
\begin{tikzpicture}
\draw (0, 0) node[inner sep=0] {\raisebox{0.1cm}{\includegraphics[height=7.5truecm,width=10.0truecm,trim={0.0cm 1.9cm 6.0cm 0.0cm}, clip]{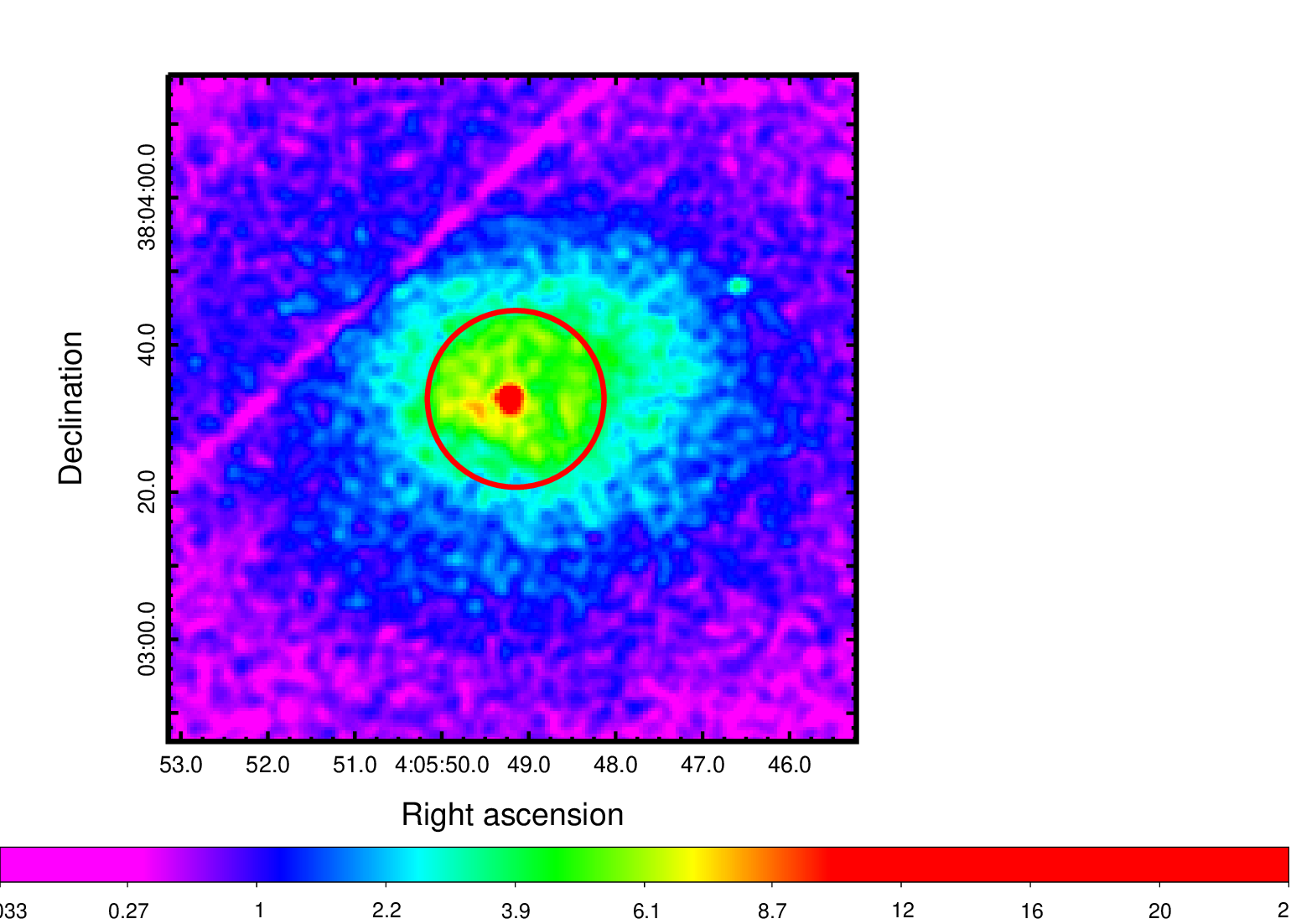}}};
\end{tikzpicture}
\hspace{-0.5cm}
\begin{tikzpicture}
\draw (0, 0) node[inner sep=0] {\raisebox{0.1cm}{\includegraphics[height=7.2truecm,width=9.0truecm,trim={0.5cm 0.0cm 0.0cm 0.0cm}, clip]{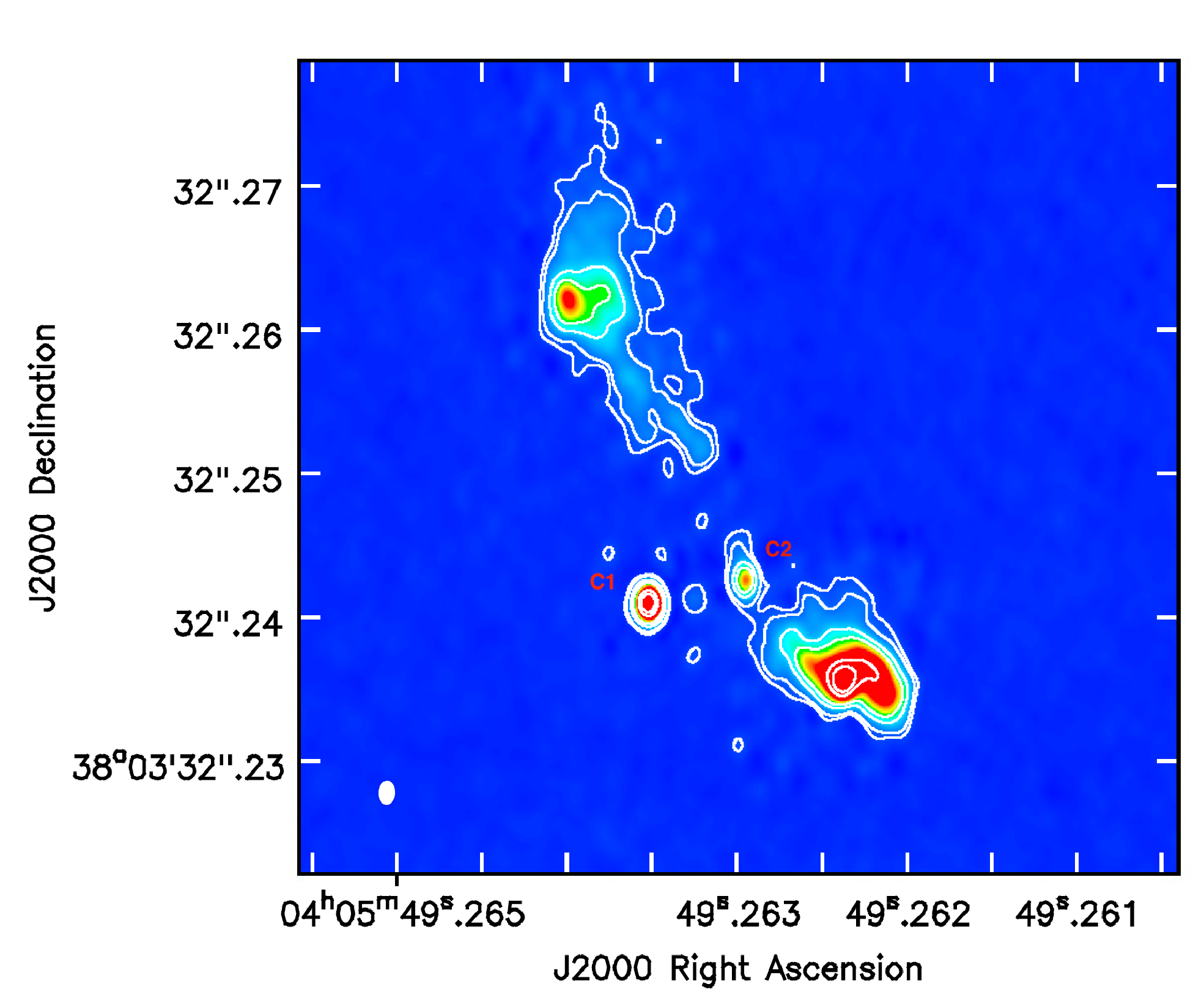}}};
\end{tikzpicture}
\hspace{-6.4cm}
\caption{{\bf Left:}The {\it Chandra} image of the core of 4C+37.11. The red circle is the region of 12$^{\prime\prime}$ considered for spectral analysis. {\bf Right:} The naturally weighted VLBA map of the zoomed-in central core region at 8 GHz where the beam size along the jet direction is 1.26 mas \citep{BansalEtal2017ApJ...843...14B}. The central core components are designated by C1 and C2 with negative contours in magenta color. The contour levels are 0.6, 1.25, 5, 10, 40, 60\% of the peak flux density 0.0792 Jy/beam.} \label{fig:ImageChandra}} 
\end{figure*}

So far, $<$200 DAGN systems have been detected  \citep{RubinurEtal2018JApA...39....8R,bhattacharya.etal.2023}. Only a few of them (e.g. 4C+37.11, NGC7674) are separated on parsec scales and have been detected with high-resolution Very Long Baseline Array (VLBA) observations \citep[][]{RodriguezEtal2006ApJ...646...49R,kharb.lal.etal.2017}. However, a recent study has claimed non-detection of the radio core in NGC 7674 \citep{BreidingEtal2022ApJ...933..143B} suggesting a further in-depth analysis of the binary AGN system. Instead, most AGN pairs have been identified at kiloparsec (kpc) scale separations (DAGN), where the two nuclei may not be bound by mutual gravitational effects \citep{OwenEtal1985ApJ...294L..85O,KomossaEtal2003ApJ...582L..15K,bhattacharya.etal.2023}. Of the binary AGN detected, 4C+37.11 is the most well-studied example. It is a radio galaxy at redshift z=0.055, harboring a binary AGN system, where the two cores (designated as C1 and C2 in Figure \ref{fig:ImageChandra}) are separated by $\sim$7 pc \citep{ManessEtal2004ApJ...602..123M,RodriguezEtal2006ApJ...646...49R}. A prominent jet (C2 in Figure \ref{fig:ImageChandra}) is also observed from one of the nuclei in the Very Long Baseline Array (VLBA) radio image \citep{BansalEtal2017ApJ...843...14B}.

Depending on the separation between the AGN, the mass accretion may change, as the gravitational potential of one affects the other. Therefore, it is quite natural to see effects in the spectral properties. Among the several complex features in X-ray AGN spectra, the presence of the Fe K$\alpha$ line and its shape is very important as it can provide valuable information regarding the accretion disk geometry, spin parameters, and physical processes occurring close to the SMBH. It is believed that the power-law spectrum in the hard X-ray band becomes modified by emission and absorption processes, that arise due to the reprocessing of radiation by cold material in the accretion disk or from more distant material such as that in the torus \citep[][and references therein]{MagdziarzZdzi1995MNRAS.273..837M,ChakrabartiTitarchuk1995ApJ...455..623C,NandraEtalPEXMON2007MNRAS.382..194N,GarciaEtal2013ApJ...768..146G}. One of the signatures of such reprocessing is the presence of the Fe K$\alpha$ line. Therefore, this line carries information about the dynamical state and thermodynamics of the reprocessing material. For example, if the disk moves much closer and reprocesses the radiation, it should be reflected in the shape of the Fe K$\alpha$, namely e.g. double peaked due to rotation and asymmetric owing to relativistic effects \citep[see e.g.][and references therein]{Fabianetal1989,Iwasawaetal1996,BrennemanRey2006ApJ...652.1028B,Mondaletal2016}. A recent study of DAGN using {\it Chandra} reported strong neutral lines from both nuclei in the merger remnant NGC 6240 \citep{Nardini2017MNRAS.471.3483N}.

Though the different types of AGN show different features in their Fe K$\alpha$ line, the width of the Fe K$\alpha$ can be a valuable probe of the emission region. Typically, equivalent widths (EW) $\sim1$ keV are associated with absorbed sources such as distant material or a torus, while EW $\sim 0.1$ keV point to unabsorbed sources such as accretion disks or broad line regions (BLR). The broad iron lines, and the associated Compton reflection continuum, are generally weak or absent in the radio-loud counterparts \citep{EracleousEtal1996ApJ...459...89E,ReynoldsBegelman1997ApJ...487L.135R,SambrunaEtal1999ApJ...526...60S,GrandiEtal1999A&A...343...33G}. This can be due to the washing out of a typical `Seyfert-like' X-ray spectrum by a Doppler-boosted jet component, or the inner disk may be unable to produce reflection features because it is in a very hot state. At the same time, it has been found that a broad component of the Fe K$\alpha$ line is present in many radio quiet (RQ; ``non-jetted") AGNs \citep{NandraEtal1997ApJ...488L..91N,PatrickEtal2012MNRAS.426.2522P,MantovaniEtal2016MNRAS.458.4198M}. Also, it is important to note that the broad line profiles help to constrain the spin parameter of the BH \citep[for a recent review see][]{ReynoldsReview2019NatAs...3...41R}.

A systematic analysis of a sample of 149 RQ type-1 AGNs by \citet{delaCalleEtal2010A&A...524A..50D} using {\it XMM–Newton} data found that 36\% of the sources show strong evidence of a relativistic Fe K$\alpha$ line. The average equivalent width (EW) of them is of the order of 100 eV. Therefore, it is still an open question whether a broad Fe line is commonly present in RL or in RQ AGNs. Earlier studies have reported the identification of a broad Fe K$\alpha$ line in a few individual RL AGNs \citep{LohfinkEtal2015ApJ...814...24L}. However, very few of them have robust spin measurements. Similarly, the narrow Fe K$\alpha$ lines observed in X-ray spectra are interpreted as fluorescence lines from neutral (cold) matter away from the inner accretion disk. There are suggestions on the origin of the narrow emission line, which include the molecular torus, the BLR, or the very outermost regions of the accretion disk \citep{JIangEtal2006ApJ...644..725J}. 

Hence, independent of the merging process and the formation of DAGN, one can ask: does the spectrum of closely separated DAGN or a binary AGN system appear similar to that of an isolated AGN disk spectrum? Can we make out if only one AGN is contributing to the observed spectrum, while the other is not very active in radio emission,  which is similar to the case of 4C+37.11? In this paper, we have tried to answer these questions by studying the X-ray spectrum of the binary AGN system 4C+37.11 using {\it Chandra} archival data. In the next section (\S 2), we discuss the observation details and data analysis. We discuss our results and findings in \S 3 and finally present our conclusions.

\section{Observation and Data Analysis}
We use archival {\it Chandra} observations (ids: 16120, 12704; hereafter O1 and O2) of 4C+37.11 observed on 06-11-2013 and 04-04-2011. The exposure times of the observations are $\sim 95$ ks and $\sim11$ ks respectively. The Chandra data are reprocessed with CIAO (v.4.14) using the updated calibration files (v.4.9.6). The level 2 event files were recreated by the script of {\tt chandra\_repro}. The source was extracted from the circular region with a radius of 12$^{\prime \prime}$ and the background with a radius of 30$^{\prime \prime}$. The {\it Chandra} image of the source and the corresponding source region in red-circle are shown in the left panel of Figure \ref{fig:ImageChandra}. The right panel of Figure \ref{fig:ImageChandra} shows the VLBA image of the zoomed-in central core region of 4C+37.11 at 8 GHz. The core components are C1 and C2 where C1 appears to be a compact and C2 shows a prominent jet emanating from the central AGN \citep{BansalEtal2017ApJ...843...14B}. The spectra of source and background, as well as the response files, were produced by the script of {\tt specextract}.  We could not find significant pile-up\footnote{https://cxc.harvard.edu/toolkit/pimms.jsp} ($<7\%$) for both spectra. The spectrum was binned using the {\tt grppha} tool to
ensure 10 counts per spectral bin before exporting to {\tt xspec} for
analysis \citep{ReynoldaMiller2010ApJ...723.1799R}. The spectral analysis is done for the energy range 0.7$-$8 keV. We tried a different source region of 6$^{\prime \prime}$, which does not change the spectral shape, however, lowers the number of data points in the whole spectrum. Different models available in {\tt XSPEC} (v.12.12.0) are used to fit the data of both observations. The spin parameter ($a$) is estimated directly from the \krd\, spectral model \citep{BrennemanRey2006ApJ...652.1028B} fitting. These model fits give accretion properties, radiation processes, and most importantly BH's intrinsic parameter (e.g. spin). All errors are estimated in 1$\sigma$ level confidence throughout the paper using {\tt err} task.

\begin{figure*}
\hspace{-1.0cm}
\begin{tikzpicture}
\draw (0, 0) node[inner sep=0] {\raisebox{0.0cm}{\includegraphics[height=6.6truecm,width=4.2truecm,angle=270,trim={0.5cm 0.0cm 0.0cm 0.0cm}, clip]{Figures/zgauss-apec-O1-r1.eps}}};
\end{tikzpicture}
\hspace{-0.3cm}
\begin{tikzpicture}
\draw (0, 0) node[inner sep=0] {\raisebox{0.1cm}{\includegraphics[height=6.5truecm,width=4.2truecm,angle=270,trim={0.5cm 0cm 0cm 2cm}, clip]{Figures/contour-zgauss-apec-O1-r1.eps}}};
\end{tikzpicture}
\hspace{-0.3cm}
\begin{tikzpicture}
\draw (0, 0) node[inner sep=0] {\raisebox{0.1cm}{\includegraphics[height=4.3truecm,width=6.5truecm,trim={1.0cm 0cm 0cm 0.0cm}, clip]{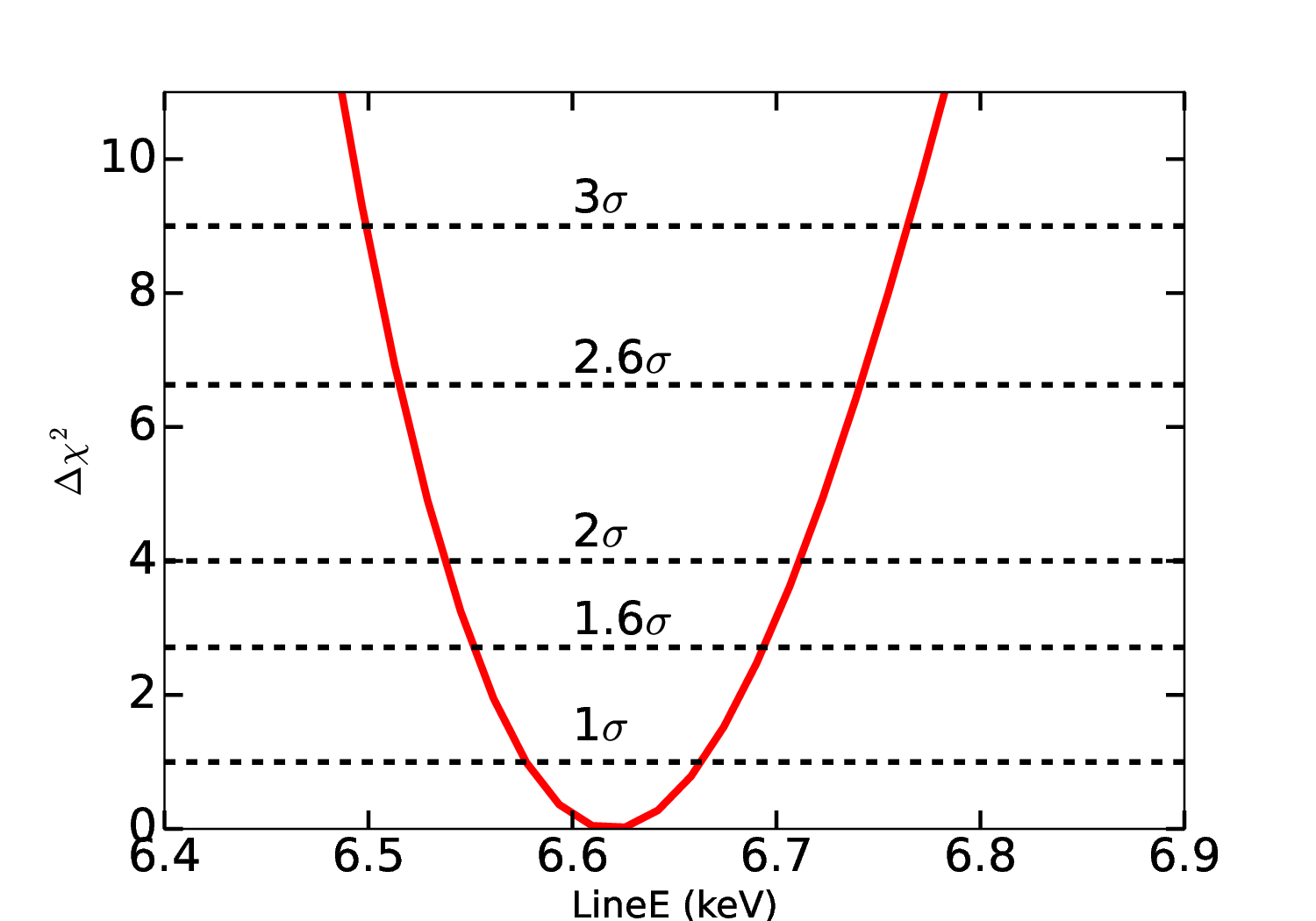}}};
\end{tikzpicture}
\hspace{-2.0cm}
\caption{{\bf Left:} M1 model fitted spectrum  of binary AGN 4C+37.11 in the 0.7$-$7.5 keV energy range. The line width of the Fe K$\alpha$ line is $\sim0.19\pm0.05$ keV. {\bf Middle:} The confidence contour between the line energy and line width. The red, green, and blue contours correspond 1$\sigma$, 2$\sigma$, and 3$\sigma$ levels,  and {\bf Right:} $\Delta\chi^2$ confidence contour of the Fe K$\alpha$ line energy. The dashed horizontal lines represent different confidence levels of the detection.}
\label{fig:SpecZgauss}
\end{figure*} 

\section{Results and Discussion}
First, we fit the O1 data using a simple power-law (\pow)\, model to fit the continuum, which returns a statistically poor fit with $\chi^2 =$ 583 for 293 degrees of freedom (dof) and the photon index ($\Gamma$) $2.59\pm0.04$. Then we perform the same fitting by adding a \zgauss\, component to the fit, to detect the Fe K line. This improves the fit with  $\chi^2 =$ 279 for 290 dof and $\Gamma=2.75\pm0.05$. The Fe K line fits at energy $\sim6.66\pm0.02$ keV with $\sigma_E \sim 82.7\pm26.5$ eV. In general, Fe K$\alpha$ emission is often observed at 6.4 keV. However, in the 6-7 keV band other Fe K lines can also be found, for instance, Fe XXV at 6.63-6.70 keV and Fe XXVI at 6.97 keV, respectively \citep[see,][and references therein]{PontiEtal2009MNRAS.394.1487P,PatrickEtal2012MNRAS.426.2522P}. Therefore, from now on, we refer to the detected line as the Fe K line. We further run {\tt ftest} task in {\tt xspec} to check the requirement of the \zgauss\,component, which yields F-statistic value = 20.98 and probability $2.47\times 10^{-12}$, that corresponds to $6.9\sigma$ confidence level. 
The O2 observation is fitted using the same model and detects the presence of the Fe K line at $6.66\pm0.11$ keV, however, due to less number of data points, we could not constrain the line width ($63^{+119}_{-61}$ eV). The model fitting with ($\chi^2 =$ 81 for 81 dof) or without ($\chi^2 =$ 86 for 84 dof) adding \zgauss\, component does not improve much, which can be due to less exposure time. The F statistic value is 1.67 and probability 0.18, which corresponds to $0.9\sigma$ level. As the Fe K line is clearly visible in our spectrum, we adopted the F-test method for the detection level of the emission line in O1, following approaches in the literature \citep[see][and others]{PontiEtal2009MNRAS.394.1487P,TombesiEtal2010A&A...521A..57T,LiuEtal2010ApJ...710.1228L,HuEtal2019MNRAS.488.4378H,LiuEtal2023arXiv231216354L}. 

Additionally, the spectrum (specially in O1) also showed soft excess below 2 keV \citep[see also,][]{ArnaudEtal1985MNRAS.217..105A,SinghEtal1985ApJ...297..633S}. For the soft excess and the presence of optically-thin collisionally ionized plasma, the \apec\, model is used, while other additive models are used to take into account the Fe K line. To obtain more insight into the system and the origin of the line, two different classes of models have been used. The first set is a single-continuum models: M1: \ztbabs(\apec+\zgauss), M2: \tbabs(\apec+\laor), M3: \tbabs(\apec+\krd), and M4: \ztbabs(\xillver+\zgauss), and the second set is two-continuum: M5: \ztbabs(\apec+\apec) and M6: \ztbabs(\relxil+\apec). These models cover both the contribution of ionized plasma due to shocked gas or the star formation activity in the host galaxy, distant reflection, and relativistic effects on the origin of the Fe K line. We first discuss the results using the analysis from the first set of models and then the second set and infer the results using the later models.

\subsection{Analysis with Single-continuum Models}
We start the model fitting using a single component \apec\,model, which could not fit the entire line in O1 observation as $\chi^2 = 357$ for 290 dof. Therefore, we fit the O1 spectrum using the M1 model, yielding a satisfactory fit with $\chi^2 =$ 330 for the dof 287. This model helps to determine the Fe K line properties (e.g. width and energy) qualitatively. The Fe K line fits at $\sim 6.62\pm0.06$ keV with the $\sigma_g$ $\sim 0.19\pm0.05$ keV for O1, while the O2 spectrum does not require \zgauss\,component. The EW of the line in O1 estimated $\sim0.36$ keV using {\tt eqwidth} task. Based on the $\sigma_g$ value, the line can be considered as a broad Fe K line. The O2 spectrum fits well with M1 (without \zgauss\,component) model provides k$T_e$ = $3.62^{+0.48}_{-0.41}$ keV and $Z = 1.14^{+0.41}_{-0.39}$ Z$_\odot$. The hydrogen column density $N_H$ obtained from the M1 model is $1.09^{+0.04}_{-0.04}\times 10^{22}$ cm$^{-2}$ (for O1) and $0.71^{+0.11}_{-0.10}\times 10^{22}$ cm$^{-2}$ (for O2) using \ztbabs\,model \citep{Wilms2000}, which takes into account the X-ray absorption through the interstellar medium (ISM). The model fitted parameters are given in \autoref{tab:SpecFit}.
The M1 model fitted spectrum for O1 is shown in the left panel of \autoref{fig:SpecZgauss}. Additionally, we have drawn the confidence contour between the line energy and line width, and the $\Delta\chi^2$ contour of the line energy. The contours are shown in the middle and right panels of \autoref{fig:SpecZgauss}. The red, green, and blue lines in the middle panel correspond to the 1$\sigma$, $2\sigma$, and $3\sigma$ confidence respectively. The dashed horizontal lines in the right panel correspond to different significance levels of the uncertainty of the position of the Fe K line. Since the Fe K line is broad and can originate from different regions of the SMBH vicinity, we first tested emission line models with relativistic effects.

During M2 model fitting to O1 observation, we varied both the \laor\, \citep{Laor1991ApJ...376...90L} and \apec\,model parameters. During this fitting, we kept the outer radius of the disk $R_{\rm out}$ fixed to 400 $r_g$ $(=GM_{BH}/c^2)$. The best-fitted emissivity index ($q$) and the disk inclination ($i$) obtained from the fit are $1.7^{+0.3}_{-0.5}$ and $86^{+1}_{-12}$. The inner edge of the disk ($R_{\rm in}$) is pegged at the lower limit, 1.24 $r_g$. The \apec\,model parameters obtained from the fit are consistent with M1 model parameters.

To further get an initial qualitative estimate of the BH spin parameter, we have fitted the data using model M3, where a similar procedure as in M2 is followed. For the M3 fitting, we varied some of the \krd\, \citep{BrennemanRey2006ApJ...652.1028B} model parameters. During this fitting, we have frozen the break radius ($R_{\rm br}$) separating the inner ($R_{\rm in}$) and outer radii of the disk ($R_{\rm out}$) to $\sim 10$ r$_g$, $i$ to $86^\circ$, and $R_{\rm in}$ to 1.24 $r_g$ obtained from M2 fitting, keeping \apec\,model parameters free. The dimensionless spin parameter obtained from the best fit is $a_k = 0.25^{+0.54}_{-0.06}$ and the emissivity index ($q_2^k$) = $1.5^{+0.2}_{-0.6}$. Since $R_{in}$ could not be constrained from M2 model and the $a_k$ parameter in M3 is providing a large error bar, therefore, the spin parameter ($a_k$) can not be tightly constrained which can have a large degeneracy \citep[see][]{FabianEtal2014MNRAS.439.2307F}. The \apec\,model parameter values obtained from the fit are consistent with M1 and M2. We note that as the Fe K line is not very significant in O2 observation due to low statistics, we could not use M2 and M3 to constrain model parameters. 
All model-fitted results are provided in Table \ref{tab:SpecFit}. 

\begin{figure*}
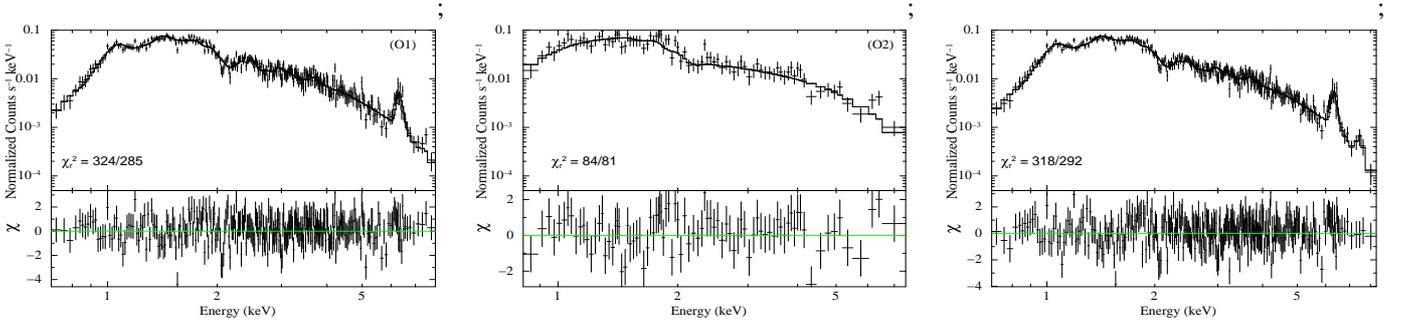

\hspace{-1.0cm}
\centering
\includegraphics[height=6.0truecm,width=4.0truecm,angle=270,trim={0.5cm 0.0cm 0.0cm 0.0cm}, clip]{Figures/R2-O1-relxill.eps};
\includegraphics[height=6.0truecm,width=4.0truecm,angle=270,trim={0.5cm 0cm 0cm 0.0cm}, clip]{Figures/R2-O2-relxill.eps};
\includegraphics[height=6.0truecm,width=4.0truecm,angle=270,trim={0.5cm 0cm 0cm 0.0cm}, clip]{Figures/R2-O12-relxill.eps};
\caption{M6 model fitted the spectra of binary AGN 4C+37.11 in the 0.7$-$8 keV energy range. The left panel shows the best fit for observation O1, and the middle panel for O2. The right panel shows the best-fitted stacked spectrum. The lower panels show the corresponding residuals of the fits.}
\label{fig:SpecMod2}
\end{figure*} 

In the next step of the analysis, we studied the possibility of the Fe line originating from a distant reflector, mainly from the toroidal region using \xillver\,model \citep{GarciaEtal2013ApJ...768..146G}. As the \apec\,model can also partly contribute to the emission line, we used \xillver\,as a standalone case without considering \apec\,component. During M4 model fitting, we let \xillver\,model parameters vary. An independent non-relativistic \xillver\,model can fit the Fe K line. However, the overall spectrum fits with $\chi^2/dof = 367/285 \sim 1.3$, which is a relatively poor fit compared to other models. The best fit to O1 observation returned $\Gamma$=$3.13^{+0.03}_{-0.02}$, Fe abundance ($A_{Fe}$) = $2.94^{+0.31}_{-0.26}$ $Z_\odot$, ionization parameter ($\log \xi$) = $3.65^{+0.13}_{-0.08}$, the electron temperature of the corona ($kT_e$) is pegged at 400 keV, and disk inclination ($i$) = $87^{\circ{+0}}_{-13}$. Since we used \xillver\,as a single-component model without \apec\,model, achieving the best fit required a \zgauss\,component at $\sim0.8$ keV to fit the excess emission around that energy. The O2 data fitting using M4 model returned $\Gamma=1.53^{+0.11}_{-0.13}$, $A_{Fe}=3.14^{+1.19}_{-1.32}$ $Z_\odot$, $\log\xi=3.16^{+0.56}_{-0.17}$, $kT_e=5.5^{+0.9}_{-1.3}$ keV, and $i = 78^{\circ{+10}}_{-4}$. In these two epochs, all model parameters remained more or less constant within the error bar in a timescale of two and a half years except $N_H$, $\Gamma$, and k$T_e$. 

\subsection{Analysis with Two-continuum Models}

The models M1-M4 give some initial qualitative information on the properties of the Fe K line. None of these models describes the realistic emission from an accretion disk. Therefore, the question remains, whether the Fe K lines originate from the accretion disk or the material unaffected by strong dynamical effects? To address that, we further used the second set of two-continuum \apec\,and relativistic reflection \relxil\,models \citep{DauserEtal2014MNRAS.444L.100D,GarciaEtal2014ApJ...782...76G}.
Since, a single-component \apec\,model alone can not fit the data well, we have added another \apec\,component (M5), which can fit the O1 spectrum well with $\chi^2 = 329$ for 287 dof. The temperature and abundances of the ionized mediums in \apec\,model components in M5 are $0.82^{+0.12}_{-0.22}$ keV and $2.57^{+0.11}_{-0.12}$ keV, and $0.14^{+0.07}_{-0.06} Z_\odot$ and $2.17^{+0.35}_{-0.30} Z_\odot$ respectively. The M5 model can successfully take into account the Fe K line as well. The $N_H$ obtained from the M5 model is $1.45^{+0.12}_{-0.09}$ $\times 10^{22}$ cm$^{-2}$. Therefore, the the origin of Fe K line can be due to multi-component optically thin collisionally ionized plasma. For the O2 spectral fitting, we did not require the additional \apec\,component.

Later, we used the relativistic reflection-based \relxil\,model (M6), which takes into account accretion disk emissions more realistically and comprehensively. This model uses an empirical broken power law emissivity and a
\texttt{cutoffpl} spectral shape for the central source. The \relxil\,model has fourteen parameters out of which some parameters can be frozen to reasonable values to avoid degeneracy in best-fitted parameters as was done by several authors in the literature \citep[see][and references therein]{InabaEtal2022ApJ...939...88I,MondalEtal2024arXiv240314169M}. From the rest of the parameters, we choose non-rotating BH as keeping it free does not allow us to constrain its value since the spin parameter ($a$) is pegging to its upper hard limit of 0.998. That also affects the estimation of other parameters. Additionally, from the M3 model as well, we could not constrain the $a$ parameter. Such tasks prompted us to set $a$ to 0 in M6. 
As the following \relxil\,model parameters: the inner edge of the disk ($R_{\rm in}$), $E_{\rm cut}$, reflection fraction ($R_{\rm ref}$), and emissivity indexes ({\it index1} and {\it index2}) cannot be constrained given the statistical quality and unavailability of the broadband data, we have fixed them to astrophysically-motivated values and continued the modeling in the next step of the analysis.

The inner and outer edges of the disk, $R_{\rm in}$ and $R_{\rm out}$ are set to $R_{\rm ISCO}$ and 1000 $r_g$. The parameter $R_{br}$ determines a break radius that separates two regimes with different emissivity profiles and is set to 10 $r_g$. We fixed the emissivity indexes at 2.4 and the reflection fraction $R_{\rm ref}$ at -0.25 to obtain only the reflected component. The above choices are made following the previous DAGN analysis by \citet[][and references therein]{InabaEtal2022ApJ...939...88I}. The $kT_e$ is set fixed to values obtained from \xillver\,model (M4) fitting. The remaining parameters, $i$, $\Gamma$, and $\log\xi$ are allowed to vary. We note that fitting the spectrum using \relxil\,model with these educated guesses could not produce the observed Fe K line, therefore, we have added an \apec\,component. The combined model reads in \texttt{XSPEC} as M6. The model fit returns an acceptable fit with $\chi^2/dof = 324/285$. The best-fitted model for O1 spectrum returns $i = 82^{+2}_{-6}$, $\Gamma = 2.52^{+0.24}_{-0.16}$, $\log\xi = 2.69^{+0.19}_{-0.24}$, and $A_{Fe}=2.85^{+0.66}_{-0.72}$ $Z_\odot$. The O2 spectral fitting gives $i = 86^{+0}_{-17}$, $\Gamma = 1.46^{+0.10}_{-0.11}$, $\log\xi = 3.28^{+0.68}_{-0.39}$ and $A_{Fe}$ pegged at the lower limit of 0.5 $Z_\odot$. Such a significantly low value of $A_{Fe}$ compared to O1 can be due to the limited number of data points in the spectrum. To further confirm this, we have redone the fit by freezing the $A_{Fe}$ value to O1, which returned a comparable fit in terms of $\chi^2$ and model parameter values.

The M6 model fitted $\xi$ is high for both O1 and O2 which is expected as the Fe K line originates from a region containing two accreting SMBHs and one of the AGN has radio jets, such regions may contain highly ionized plasma. Our estimated line energy follows the estimation of $\xi$ as discussed in the literature \citep{KallmanEtal2004ApJS..155..675K,PontiEtal2009MNRAS.394.1487P,PatrickEtal2012MNRAS.426.2522P}. The estimated absorbed total fluxes using M6 in the energy band 0.7-8.0 keV are 1.12 and 1.28 $\times 10^{-12}$ ergs cm$^{-2}$ s$^{-1}$ for the epochs O1 and O2. All parameters from M6 model fit are given in \autoref{tab:SpecFit} and the best-fitted M6 spectra are shown in the left and middle panel of \autoref{fig:SpecMod2}. Since both M5 and M6 models can fit the Fe K line satisfactorily, to quantify their contribution, we have calculated the relative fraction of the line intensity in the 6.0-7.0 keV band due to \apec\,and \relxil\,in M6. If $I_\apec$ and $I_\relxil$ are the line intensities from each model, then the relative fractions due to \apec\,model and \relxil\,models are $I_\apec$/($I_\apec$+$I_\relxil$) $\sim 60\%$ and $I_\relxil$/($I_\apec$+$I_\relxil$) $\sim 40\%$ respectively. Therefore, collisionally ionized plasma contributes more to the Fe K line.

\begin{table}
    \centering
\scriptsize    
\caption{\label{tab:SpecFit} Best fitted M1-M6 spectral model parameters of 4C+37.11. Here, $^f$ and $^p$ denote the frozen parameters and pegged to the upper hard limits of the parameters.  The $^l$ and $^k$ denote \laor\, and \krd\, model parameters respectively. $N_H$ and $\sigma_g$ denote the neutral hydrogen column density in \ztbabs\, model and Fe K line width in \zgauss\, model. See text for the details. }
\begin{tabular}{ccccccccc}
\hline
& & &\multicolumn{2}{c}{Observations}\\
\cline{4-5}

\multicolumn{2}{c}{Models}&Parameters&O1&O2\\
\hline
\multicolumn{2}{c}{\bf Single-continuum}\\
\multirow{3}{*}{M1}&\ztbabs   &$N_H$ ($10^{22}$ cm$^{-2}$) &$1.09^{+0.04}_{-0.04}$&$0.71^{+0.11}_{-0.10}$\\
&\zgauss                   &$E_g$ (keV)&$6.62^{+0.06}_{-0.06}$&-\\
&                           &$\sigma_g$ (keV)&$0.19^{+0.05}_{-0.05}$&-\\		
&\apec                     &k$T_e$ (keV)&$2.21^{+0.07}_{-0.07}$ &$3.62^{+0.48}_{-0.41}$\\
&                           &$Z(Z_\odot)$&$0.90^{+0.14}_{-0.12}$ &$1.14^{+0.41}_{-0.35}$\\
&Fit Statistics&$\chi^2/dof$  &330/287&71/83\\ 
\hline
\multirow{4}{*}{M2}&\ztbabs &$N_H$ ($10^{22}$ cm$^{-2}$)&$1.16^{+0.05}_{-0.04}$&\multirow{7}{*}{-}\\
&\apec                     &k$T_e$ (keV)&$2.23^{+0.08}_{-0.08}$ &\\
&                          &$Z(Z_\odot)$&$1.01^{+0.17}_{-0.13}$ &\\
&\laor                     &$E_l$&$6.5^{+0.1}_{-0.1}$&\\
&                           &$q^l$&$1.7^{+0.3}_{-0.5}$&\\
&                           &$R_{in}^l (r_g)$&$1.24^p$&\\
&                           &$i (^\circ)$&$86^{+1}_{-12}$& \\
&Fit Statistics&$\chi^2/dof$&343/287&\\ 
\hline
\multirow{4}{*}{M3}&\ztbabs &$N_H$ ($10^{22}$ cm$^{-2}$)&$1.16^{+0.05}_{-0.05}$&\multirow{7}{*}{-}\\
&\apec                     &k$T_e$ (keV)&$2.15^{+0.07}_{-0.07}$ &\\
&                          &$Z(Z_\odot)$&$0.91^{+0.12}_{-0.12}$ &\\
&\krd                      &$q_1^k$&$1.7^f$&\\
&                          &$q_2^k$&$1.5^{+0.2}_{-0.6}$&\\
&                           &$a_k$&$0.25^{+0.54}_{-0.06}$&\\
&Fit Statistics&$\chi^2/dof$&334/287&\\ 
\hline
\multirow{4}{*}{M4}&\ztbabs &$N_H$ ($10^{22}$ cm$^{-2}$)&$1.94^{+0.16}_{-0.10}$&$0.65^{+0.14}_{-0.13}$\\
&\zgauss                    &$E_g$ (keV)&$0.8^f$ &-\\
&                           &$\sigma_g$ (keV)&$0.17^{+0.04}_{-0.03}$&-\\
&\xillver                   &$\Gamma$&$3.13^{+0.03}_{-0.02}$&$1.53^{+0.11}_{-0.13}$\\
&                           &$A_{\rm Fe}(Z_\odot)$&$2.94^{+0.31}_{-0.26}$&$3.14^{+1.19}_{-1.32}$\\
&                           &$\log\xi$&$3.65^{+0.13}_{-0.08}$&$3.16^{+0.56}_{-0.17}$\\
&                           &k$T_e$ (keV)&$400^p$&$5.5^{+0.9}_{-1.3}$\\
&                           &$i(^\circ)$&$87^{+0}_{-13}$&$78^{+10}_{-4}$\\
&Fit Statistics&$\chi^2/dof$&367/285&78/80\\ 
\hline
\multicolumn{2}{c}{\bf Two-continuum}\\
\multirow{3}{*}{M5}&\ztbabs   &$N_H$ ($10^{22}$ cm$^{-2}$) &$1.45^{+0.12}_{-0.09}$&\multirow{6}{*}{-}\\
&\apec1                     &k$T_{e1}$ (keV)&$0.82^{+0.12}_{-0.22}$&\\
&                           &$Z_1(Z_\odot)$&$0.14^{+0.07}_{-0.06}$&\\
&\apec2                     &k$T_{e2}$ (keV)&$2.57^{+0.11}_{-0.12}$&\\
&                           &$Z_2(Z_\odot)$&$2.17^{+0.35}_{-0.30}$ &\\
&Fit Statistics&$\chi^2/dof$  &329/287&\\ 
\hline
\multirow{4}{*}{M6}&\ztbabs       &$N_H$($10^{22}$ cm$^{-2}$)&$1.43^{+0.06}_{-0.08}$&$0.64^{+0.16}_{-0.10}$\\
&\apec                 &k$T_e$ (keV)&$2.28^{+0.08}_{-0.11}$ &-\\
&                      &$Z(Z_\odot)$&$2.12^{+0.39}_{-0.32}$ &-\\
&\relxil            &$i$ $(^\circ)$&$82^{+2}_{-6}$&$86^{+0}_{-17}$\\
&			        &$\Gamma$&$2.52^{+0.24}_{-0.16}$&$1.46^{+0.10}_{-0.11}$\\
&			   &$\log\xi$ &$2.69^{+0.19}_{-0.24}$&$3.28^{+0.68}_{-0.39}$\\
&			   &$A_{\rm Fe}(Z_\odot)$&$2.85^{+0.66}_{-0.72}$&$0.5^p$\\
&			   &$E_{\rm cut}$ (keV)&$400^f$&$5.5^f$\\
&Fit Statistics&$\chi^2/dof$  &324/285&84/81\\     
\hline    
\end{tabular} 
\end{table}

During M1-M6 model fits to O1 spectrum showed another line feature above 7.5 keV (see left panels of \autoref{fig:SpecZgauss} and \autoref{fig:SpecMod2}). However, the data statistic at that line energy is poor. Therefore, we stacked both spectra, which shows a clear line feature at higher energy. We have fitted this line component by adding another \zgauss\,component to M1 to find the line properties, which gives $E_g$ $\sim 7.87^{+0.09}_{-0.11}$ keV with $\sigma_g\sim0.21^{+0.19}_{-0.09}$ keV, that keeps the \apec\, model parameters unchanged. Further, we have fitted the stacked spectrum using M6 model by adding another \zgauss\,component for the second Fe K line, which gives $E_g$ $\sim 7.87^{+0.11}_{-0.09}$ keV with $\sigma_g\sim0.07^{+0.18}_{-0.06}$ keV, with k$T_e=2.32^{+0.10}_{-0.12}$ keV and $Z=1.81^{+0.33}_{-0.29} Z_\odot$ in \apec\, model. The other \relxil\,model parameters are $\Gamma=2.49^{+0.22}_{-0.14}$, $A_{Fe}=2.75^{+0.70}_{-0.74}$, $\log\xi=2.59^{+0.25}_{-0.18}$, $i=84^{\circ+3}_{-8}$, and $kT_e$ was frozen at 400 keV. The model fits the data very well with $\chi^2 =317$ for 291 dof. The best-fitted model spectrum is shown at the right panel of \autoref{fig:SpecMod2}. Such line feature $\sim 7.8-7.9$ keV can be the emission line due to Fe K$\beta$ or Ni K$\alpha$ or the mixture of both coming from a distant material. The difference in peak line energies also indicates that those may not be coming from the individual nuclei, as the relative velocity required to obtain such a large energy difference ($\Delta E>1.2$ keV $\sim0.19c$) exceeds the escape velocity of the BSMBH system ($\sim0.01c$ for the system mass $\sim10^{10} M_\odot$ and separation 7 pc, orbiting in circular orbit). This indicates that the origin of the lines is a distant-ionized material. However, as this is a binary system, the scenario might be more complex, therefore follow-up studies are required to better constrain the Fe K lines and their origin.

The M6 model fitted $N_H$ value showed a significant change between epochs in 2.5 years, however not very high, $\leq 1.5\times10^{22}$ cm$^{-2}$. It was shown that the star formation rates (SFR) and
AGN fractions are expected to increase as a function of
decreasing pair separation of mergers \citep[e.g.,][]{EllisonEtal2008AJ....135.1877E}, so we should expect that dual or binary  AGNs generally show a gas-rich environment thereby, an elevated SFR. However, recent galaxy merger simulations showed that both SFR and circumnuclear obscuration (in terms of $N_H$) become weaker in gas-poor mergers than in gas-rich ones \citep{BlechaEtal2018MNRAS.478.3056B}, which also depends on the gas-to-mass ratio of the host galaxy \citep{InabaEtal2022ApJ...939...88I}. The galaxy 4C+37.11 is an elliptical galaxy. The gas in an elliptical galaxy is little or no star formation \citep{RomaniEtal2014ApJ...780..149R}. Hence, there is probably not much gas associated with star formation in this system. Our results favor the gas-poor merger scenario. On the other hand, the change in hot ionized gas temperatures in M1 between O1 and O2 and in \apec\,model components (M5) in O1 can be due to shocked gas associated with the binary environment, possibly not due to star formation activities.

Given the best-fitted M6 model parameters and luminosities ($L_X$ $\sim8.2$ and $9.1\times 10^{42}$ erg s$^{-1}$) for both observations, the location of the ionized medium can be derived from the definition of $\xi$ \citep{TarterEtal1969ApJ...156..943T} as $R\sim L_X/N_H\xi \sim 0.7-1.1\times10^{18}$ cm. The estimation is based on assumptions that the medium has a constant density and the thickness of the medium does not exceed its distance from the central SMBH \citep[e.g.,][]{CrenshawKraemer2012ApJ...753...75C,MesticiEtal2024MNRAS.532.3036M}. The ionized medium can be composed of outflowing gas \citep{KingPounds2003MNRAS.345..657K} arising from within the 1 pc scale that blocks the central radiation. If we assume that the broadening of the Fe K line is due to the velocity of the distant outflowing gas, we can obtain the velocity from the EW measurement using $\frac{\varv}{c}=\frac{{\it FWHM}}{E_0}$, which yields $\varv$ $\approx 0.05c$. Here, c, ${\it FWHM}$, and $E_0$ are the speed of light, full-width half maximum of the line (from M1 fit), and rest-frame energy of the Fe K line ($\sim6.6$ keV). If one equates this velocity with the escape velocity of the outflowing gas within the gravitational potential of the central SMBH at a distance $R$, then the total mass of the BSMBH comes out to be $\sim 1.2-1.7\times10^{10}$ M$_\odot$, with an average of $\sim1.5\times10^{10}$ M$_\odot$. Our estimation agrees with the estimation using radio observation \citep{BansalEtal2017ApJ...843...14B}.

\section{Conclusion}
In this paper, we studied two {\it Chandra} observations of a binary AGN 4C+37.11 observed during 2013 (obs. O1) and 2011 (obs. O2). We report the detection of the Fe K lines at $\sim 6.62^{+0.06}_{-0.06}$ keV (with $6.9\sigma$ confidence using F-test) of width $\sim 0.19^{+0.05}_{-0.05}$ keV (1$\sigma$ confidence) and another line feature at $\sim7.87^{+0.19}_{-0.09}$ keV of width $\sim0.21^{+0.19}_{-0.09}$ keV from the stacking of two spectra in the binary AGN system 4C+37.11, where the SMBHs are separated by $\sim 7$pc. Here, we briefly discuss the origin of the Fe K line from the model-fitted parameters. 

The consideration of a simple fluorescent line (model M1) fits the lines well and provides the properties of the Fe K lines. However, can not infer their origin, as the system contains two black holes, where both of them have accretion disks, one of them contains a radio jet, and the circumnuclear region is associated with ionized gas. Therefore, we used two sets of models, the first one is a single-continuum model (M1-M4) and the second one is a two-continuum model (M5-M6). The former sets of models can give us preliminary information on the nature of the lines and the accretion disk emissivity properties qualitatively. It is believed that the effect of spin or the origin of the Fe K line from the very inner edge of the disk leaves its footprint by a broadening or skewing of the line profile. Since the Fe K lines in 4C+37.11 are broad and if it is due to relativistic effects, then model M3 is supposed to be able to constrain the spin parameter, however, it could not. It may support that the lines did not originate from the accretion disk and further suggests a study using models that consider the distant reflection. The M4 model can fit the Fe K line independently, however, the overall fit statistic is poor (reduced $\chi^2 \sim 1.3$). Extensive tests with single-continuum models and not having promising results prompted us to further use the two-continuum models (M5 and M6), especially the \relxil\,model which takes into account emission from an accretion disk realistically.

A two-continuum collisionally ionized plasma model (\apec, M5) can fit the observed Fe K line profile, which infers a possible origin of the line from a circumnuclear environment. The above model fits can constrain the temperature and abundance of the scattering medium, where the abundances of the material vary between sub-solar to super-solar regimes. Later, we used the M6 model that can fit the line profile as well as the continuum with acceptable fit statistics, which further favors the origin of the line profiles from the combined effects of the accretion disk and the circumnuclear environment. The high values of both Fe abundance and ionization parameters suggest that the Fe K lines possibly originated from highly ionized mediums. This is also reflected in the line energy, which is shifted by 0.2 keV compared to the neutral Fe K$\alpha$ line. Since the M6 model takes into account both emissions from the accretion disk and collisionally ionized plasma and provides additional information about the system, this model is more favored compared to M5 for the present study.
 
From the data fitted by two sets of models, we conclude the following:

\begin{enumerate}

    \item Single-continuum models along with line emission (M1-M4) could not explain the origin of Fe K lines, however, provide the initial line properties qualitatively.

    \item Both two-continuum models (M5 and M6) can represent the line properties satisfactorily, suggesting the line originated from the circumnuclear ionized medium. Therefore, infers their origin due to combined effects rather than in isolation from the accretion disk.  
    
    \item The best-fitted \relxil\,model suggests the inclination of the disk is $\gtrsim 75^\circ$ and the medium from where the line originates is highly ionized ($\log\xi\gtrsim 2.7$) with sub to super-solar Fe abundance. The total mass of the BSMBH is estimated using the M6 model parameters and luminosity to be $\sim1.5\times10^{10}$ M$_\odot$.
    
    \item A low nuclear obscuration (in terms of $N_H$ in M6) in elliptical galaxy 4C+37.11 supports the gas-poor merger scenario more, than the elevated star formation activities. Also, the significant change in hot ionized medium temperature can be associated with the shocked gas in the binary merger and not with the SFR.

    \item We could not constrain inner edge of the disk $(R_{in})$ and the spin parameter $(a)$ from the present analysis.

    \item We note that M5 and M6 are alternative scenarios. Therefore, one can derive constraints on $a$ and $R_{in}$ of the accretion flow only if the M6 model is the correct representation of the data and no statement on the same quantities can be drawn if M5 is the correct description.

\end{enumerate}

Follow-up broadband X-ray observations are needed to understand this system in detail. Furthermore, onboard and future microcalorimeters such as XRISM \citep[e.g.,][]{XRISM_SCIENCE_Team2020arXiv200304962X} and Athena \citep[e.g.,][]{BarretEtal2016SPIE.9905E..2FB} will revolutionize such estimations due to their high spectroscopic resolution, and will possibly be able to confirm this detection. For the radio observations, we have adequate resolution, but additional observations are needed to better constrain the orbital parameters of the binary black hole system. In the optical band, the H$\beta$ line detection can give further clues about the emission region of the Fe K lines \citep{Nandra2006MNRAS.368L..62N}, if such lines can be detected in the future.

\section{Acknowledgements}
We thank the referee for making thoughtful comments and suggestions that help improve the quality of the paper. SM thanks V. Liu for helping with some analysis. SM acknowledges the Ramanujan Fellowship (RJF/2020/000113) by DST-SERB, Govt. of India for this research. M.D. acknowledges the support of the Science and
Engineering Research Board (SERB) CRG grant CRG/2022/004531 for this research. AN acknowledges the Summer Research Fellowship program of IAS, INSA, NASI for providing financial support. The National Radio Astronomy Observatory is a facility of the National Science  Foundation operated under cooperative agreement by Associated Universities, Inc.  This research has made use of data obtained from the Chandra Data Archive and the Chandra Source Catalog, and software provided by the Chandra X-ray Center (CXC) in the application packages CIAO and made use of the {\it NuSTAR} Data Analysis Software ({\sc nustardas}) jointly developed by the ASI Science Data Center (ASDC), Italy and the California Institute of Technology (Caltech), USA. This research has also made use of data obtained through the High Energy Astrophysics Science Archive Research Center Online Service, provided by NASA/Goddard Space Flight Center.

\section*{Data Availability}
Data and models used for this study are publicly available on NASA's HEASARC website and in {\it XSPEC} software package.


\bibliography{reference} 
\bibliographystyle{aa}

\end{document}